\documentstyle[12pt,epsf]{article}

\newfont{\lfont}{line10}

\newcommand{\BE}{\begin{equation}}
\newcommand{\EE}{\end{equation}}
\newcommand{\BEA}{\begin{eqnarray}}
\newcommand{\EEA}{\end{eqnarray}}

\newcommand{\IJMP}[3]{Int. J. Mod. Phys. {\bf #1}{(#2)}{#3}}
\newcommand{\JHEP}[3]{JHEP~{\bf #1}{(#2)}{#3}}

\newcommand{\NP}[3]{Nucl. Phys. {\bf #1}{(#2)}{#3}}
\newcommand{\PL}[3]{Phys. Lett. {\bf #1}{(#2)}{#3}}
\newcommand{\PR}[3]{Phys. Rev. {\bf #1}{(#2)}{#3}}
\newcommand{\PRL}[3]{Phys. Rev. Lett. {\bf #1}{(#2)}{#3}}
\newcommand{\PTP}[3]{Prog. Theo. Phys. {\bf #1}{(#2)}{#3}}

\def\12{\frac{1}{2}}
\def\bea{\begin{eqnarray}}
\def\eea{\end{eqnarray}}
\def\ba{\begin{array}}
\def\ea{\end{array}}

\def\one-loop{\mbox{\scriptsize one-loop}}

\def\G{\Gamma}

\catcode`\@=11

\@addtoreset{equation}{section}
\def\theequation{\arabic{section}.\arabic{equation}}

\def\@normalsize{\@setsize\normalsize{15pt}\xiipt\@xiipt
\abovedisplayskip 14pt plus3pt minus3pt%
\belowdisplayskip \abovedisplayskip
\abovedisplayshortskip  \z@ plus3pt%
\belowdisplayshortskip  7pt plus3.5pt minus0pt}

\def\small{\@setsize\small{13.6pt}\xipt\@xipt
\abovedisplayskip 13pt plus3pt minus3pt%
\belowdisplayskip \abovedisplayskip
\abovedisplayshortskip  \z@ plus3pt%
\belowdisplayshortskip  7pt plus3.5pt minus0pt
\def\@listi{\parsep 4.5pt plus 2pt minus 1pt
            \itemsep \parsep
            \topsep 9pt plus 3pt minus 3pt}}

\def\underline#1{\relax\ifmmode\@@underline#1\else
        $\@@underline{\hbox{#1}}$\relax\fi}
\@twosidetrue

\relax

\catcode`@=12

\catcode`\@=11

\def\section{\@startsection{section}{1}{\z@}{3.5ex plus 1ex minus
   .2ex}{2.3ex plus .2ex}{\large\bf}}

\def\thesection{\Roman{section}.}

\def\appendix{\setcounter{section}{0}
        \def\thesection{Appendix }
       \def\theequation{\Alph{section}.\arabic{equation}}}

\def\ps@headings{\def\@oddfoot{}\def\@evenfoot{}
\def\@oddhead{\hbox{}\hfill
        \makebox[.5\textwidth]{\raggedright\ignorespaces --\thepage{}--
        \hfill {}}}
\def\@oddhead{\hbox{}\hfill --\thepage{}-- \hfill
        {}}
\def\@evenhead{\@oddhead}
\def\subsectionmark##1{\markboth{##1}{}}
}

\ps@headings

\catcode`\@=12

\relax

\def\figcap{\section*{Figure Captions\markboth
        {FIGURECAPTIONS}{FIGURECAPTIONS}}\list
        {Fig. \arabic{enumi}:\hfill}{\settowidth\labelwidth{Fig. 999:}
        \leftmargin\labelwidth
        \advance\leftmargin\labelsep\usecounter{enumi}}}
 \relax
\def\tablecap{\section*{Table Captions\markboth
        {TABLECAPTIONS}{TABLECAPTIONS}}\list
        {Table \arabic{enumi}:\hfill}{\settowidth\labelwidth{Table 999:}
        \leftmargin\labelwidth
        \advance\leftmargin\labelsep\usecounter{enumi}}}
 \relax
\def\reflist{\section*{References\markboth
        {REFLIST}{REFLIST}}\list
        {[\arabic{enumi}]\hfill}{\settowidth\labelwidth{[999]}
        \leftmargin\labelwidth
        \advance\leftmargin\labelsep\usecounter{enumi}}}
 \relax

\catcode`\@=11

\def\ps@headings{\def\@oddfoot{}\def\@evenfoot{}
\def\@oddhead{\hbox{}\hfill
        \makebox[.5\textwidth]{\raggedright\ignorespaces --\thepage{}--
        \hfill {}}}
\def\@evenhead{\@oddhead}
\def\subsectionmark##1{\markboth{##1}{}}
}

\ps@headings

\relax

\newskip\humongous \humongous=0pt plus 1000pt minus 1000pt

\newif\ifdtup

\def\beq{\begin{equation}}
\def\eeq{\end{equation}}

\def\beqn{\begin{eqnarray}}
\def\eeqn{\end{eqnarray}}
\relax

\def\G2{{\; \rm GeV/}c^2}
\def\G{\; \rm GeV}

\def\dotx{\dotx{\dot\overline{x}}}

\relax

\begin{document}
%
%
\begin{titlepage}

\renewcommand{\thefootnote}{\fnsymbol{footnote}}

\begin{flushright}
      \normalsize
    January, 2002  \\
     OU-HET 404, \\
     KEK-TH-794, \\
        hep-th/0201035 \\
     revised version \\
\end{flushright}

%
\begin{center}
  {\large\bf Extension of
  Boundary String Field Theory \\
  on Disc and $RP^{2}$  Worldsheet Geometries}
\footnote{This work is supported in part
 by the Grant-in-Aid  for Scientific Research
(12640272) from the Ministry of Education,
Science and Culture, Japan}
\end{center}

\vfill

\begin{center}
    {%
H.~Itoyama\footnote{e-mail: itoyama@het.phys.sci.osaka-u.ac.jp}
\quad and \quad
S.~Nakamura\footnote{e-mail: nakashin@post.kek.jp}
}\\
\end{center}

\vfill

\begin{center}
      ${}^{\dag}$\it  Department of Physics,
        Graduate School of Science, Osaka University,\\
        Toyonaka, Osaka 560-0043, Japan  \\

      ${}^{\ddag}$\it Theory Division,
      Institute of Particle and Nuclear Studies,\\
      High Energy Research Organization (KEK),
       Tsukuba, Ibaraki 305-0801, Japan

\end{center}

\vfill


\begin{abstract}

   We present a construction of open-closed string field
theory based on disc and $RP^2$ geometries.
Finding an appropriate $BRS$ operator in
the case of the $RP^{2}$ geometry, we generalize
 the background independent open string field theory (or boundary string
field theory) of Witten on a unit disc.
 The coupling constant flow at the closed string side is driven by
 the scalar operator inserted at the nontrivial loop of $RP^{2}$.
 We discuss the off-shell extension of the boundary/crosscap states.
Our construction provides an interpolation of orientifold planes  of
 various dimensions as well as that of $D$-branes.

\end{abstract}

\vfill

\setcounter{footnote}{0}
\renewcommand{\thefootnote}{\arabic{footnote}}

\end{titlepage}
\section{Introduction}

  For a long time,  a number of physicists have suspected that
  ``space of all two-dimensional field theories'', if it is  to be defined,
  should be a natural configuration space of strings on which
 string field theory is constructed  with an appropriate
 gauge invariant action. Ultraviolet divergences associated with irrelevant
operators
  carrying arbitrarily high dimensions have hampered progresses toward this
goal.
 It was shown by Witten  \cite{wittenbsft}\footnote{
 See also \cite{shatashvili}.},
 however, that, modulo this difficulty, a clear-cut
framework based on  Batalin-Vilkovisky formalism (BV) \cite{BV-1,BV-2}
is possible to find
in the case of open-string field theory. This framework materializes
    the idea of ``theory space'' on a unit disc.
 String field theory\footnote{ For a recent review on string field theory,
 see,  for example, \cite{arefeva}.} of this kind, originally called
background independent
 open string field theory, provides a self-consistent framework for
off-shell processes
 in which relevant and marginal operators are involved.

 More recently, this theory has been successfully applied to the problem of
open string tachyon condensation
 \cite{gerasimov-shatashvili,kutasov-marino-moore}.
 It provides
 an off-shell interpolation between the Neumann and the Dirichlet boundary
conditions of an open string.
An exact tachyon potential
\cite{gerasimov-shatashvili,kutasov-marino-moore}
 with the right normalization coefficient \cite{Gosen} has been obtained
 and this has provided a proof of Sen's conjecture \cite{sen} that
  the difference between the two extrema just cancels the tension of the
  $D$-brane in question.
 It allows us to consider the decay of a higher dimensional brane to a
lower one as well.
 An agreement on tensions of branes of various dimensions has been given.
See
\cite{0010108,Mori-Naka,tseytlin11/2000,0012198,0012210,0103089,0103102}
for some of the subsequent developments.
 We will refer to the background independent open string field theory of
Witten
  as boundary string field theory (BSFT), following the current
nomenclature.
 An extension of this framework to closed string field theory appears to be
formidable
 as we do not yet understand well enough processes under which the matter
central charge changes
  by the presence of BRS noninvariant operators located on the bulk of a
Riemann surface.

 One may note with this last regard that there are some
processes of
 a closed string and geometries represented by a closed string which do not
involve operators living in the entire
 bulk of a Riemann surface. In particular, orientifold planes of various
dimensions are
 characterized as crosscap states on a complex plane representing the $RP^2$
worldsheet geometry
 and can be discussed in parallel to $D$-brane boundary states representing
the disc worldsheet geometry.
 These points have prompted us to consider a $RP^2$ generalization of BSFT,
  an extension of the idea of theory space on the unit disc
and hence open-closed string field theory  from the disc and
 the $RP^2$ worldsheet geometries, which
share the same Euler number. For recent discussions on the other aspects
 of closed string geometries and closed string tachyon, see, for example,
\cite{Nakamura,suyama,Dont,Vafa,Dab-1,Localized,minwalla}.

  In this paper,  we will present a construction of
 open-closed  bosonic string field
 theory\footnote{See \cite{0105272}, for a recent discussion.  For
earlier references
 on the open -closed mixed system, see,  for example, \cite{mixed} }.
Our construction is based on the BV formalism applied both to
  the correlators on the boundary of the disc and the ones on the nontrivial
loop  ${\cal C}$ of $RP^{2}$ (${\bf \pi}_1(RP^2)={\bf Z}_2$). The former
part is
 a repetition of
 \cite{wittenbsft} except for the insertion of Chan-Paton space. As for the
latter, let us first note that
  this nontrivial loop can be identified as a source of the crosscap
contributing
to the Euler number  when $RP^2$ is represented as a part of the complex
plane.
We find that the scalar operators located on this nontrivial loop
accomplish the desirable interpolation between the orientifold planes of
various
dimensions
  in much the same way as the operators of BSFT do
  between $D$-branes of various dimensions.
 An off-shell extension \cite{Lee,Fujii-Itoyama-2}, (see also
  \cite{Akh-Laid-Seme}), of
$D$-brane boundary states and that of crosscap
states \cite{Horava,Callan-etal,Arakane-etal}
 provide an expedient tool for this.  The latter will be given in this
paper.

 We find an appropriate BRS operator which acts on the operators on this
loop on $RP^2$.
 Finding such an operator is mandatory for us to materialize the BV
formalism on $RP^2$.
 The total action resulting from our discussion is simply a sum of two
terms,
  one from the disc geometry and the other from $RP^2$.
In this sense, our construction generalizes that of  the first quantized
unoriented
 string theory at Euler number one.
 Each part of the action separately
 obeys  a first order differential equation with respect to the couplings of
operators inserted.
 The relative normalization between the two terms is fixed by appealing to
the dilaton tadpole cancellation at Euler number one
\cite{Grins-Wise,Doug-Grins,Itoyama-Moxhay}.
The work of \cite{Gosen} tells that  the normalization of the disc part is
in fact $D$-brane tension.
In our construction, closed string amplitudes are still beyond our reach and
    our discussion is essentially at the classical level.

 In the next section, we briefly summarize  essential ingredients of BSFT,
 which are  the BV formalism,  the action of the BRS operator on the
operators  located
 on the boundary of a disc and the defining differential equation for the
 action $S^{disc}$.  We recall the off-shell  boundary states introduced in
  \cite{Lee,Fujii-Itoyama-2}.
In section three,  the $RP^{2}$ extension of BSFT is given.
 We find an appropriate BRS operator acting properly on operators
 on the nontrivial loop represented as half of a unit circle in
 the complex plane. We compute the action of the BRS operator
 and derive a differential equation for $S^{RP^2}$.
  We introduce off-shell crosscap states which manage to interpolate
 states representing orientifold geometries of various dimensions.
In section four, we discuss the total action.
 The total action
 consists of the contributions from the two geometries.
  We discuss how to fix the relative normalization.

In Appendix A,  we construct a holomorphic field with weight p on the
disk and
that on $RP^{2}$, using  the construction on the plane as the double of the
surfaces.
These are found to be useful in the calculation in the text.
In Appendix B, we recall the tachyon vertex operator of an unoriented
  bosonic string with intercept $-4$.

\section{Background Independent Open String Field Theory of Witten,
and Off-Shell Boundary States}

\subsection{BV Formalism}

  Let us first recall the BV formalism briefly. The basic ingredients of
 this formalism can be summarized as a triplet
\beqn
\label{triplet}
   \left( {\cal M} \;, \; \omega \;, \; V  \right) \;\;\;.
\eeqn
  Here ${\cal M}$ is a supermanifold equipped with local coordinates
$\{ \{  u^{I} \} \}$.   We denote by $\omega$ a nondegenerate odd
(fermionic)
  symplectic two-form  which is closed
\beq
\label{omegaclosed}
  d \omega = 0 \;\;.
\eeq
  Finally, ${\cal M}$ possesses $U(1)$ symmetry generated by the fermionic
vector field
 $V = V^{I} \frac{d}{du^{I}}$. This $U(1)$ symmetry is identified with
  ghost number.

 The action functional (or zero form) $S$ in this formalism is introduced
  through
\beq
\label{sdomega}
 dS = i_{V} \omega \;\;,
\eeq
 where  $i_{V}$ is an interior contraction by $V$.  That $V$ generates a
symmetry of
$\omega$ translates into
\beq
\label{elomega}
  {\cal L}_{V} \omega = 0 \;\;\;,
\eeq
 where   ${\cal L}_{V} \equiv d i_{V} + i_{V} d$ is
 the Lie derivative.
  We see  that eqs. (\ref{omegaclosed}) and (\ref{elomega})
  ensure  the existence of a scalar functional $S$ obeying eq.
(\ref{sdomega})
 modulo global problems on ``theory space''.
 That the antibracket
\beq
\label{ss}
  \left\{ S, S \right\}_{AB}
\eeq
 be constant
 provides  the nilpotency of $V$
\beq
 V^{2}= 0 \;\;,
\eeq
  and $V$ is naturally identified with the BRS charge.
  In the next section, we will find that  this framework is realized on
 the $RP^{2}$
  worldsheet geometry  as well.
 Let us first review  the realization on the disc by Witten.

\subsection{ $\{ Q_{BRS}, {\cal O} \}$ }
 We will first obtain the $BRS$ transformation of a generic operator
$\cal{O}$ with
  ghost number one located on the boundary $\partial \Sigma$
 of the unit disc $\Sigma$.
  Let
\beq
\label{ocv}
{\cal O}\left(w, \bar{w} \right) =
  \frac{1}{i w} c^{z}(w) {\cal V}\left(w, \bar{w} \right)
=  c^{\sigma}\left( \sigma \right)
{\cal V}\left(w, \bar{w} \right)  \;\;\;,
\eeq
 where  ${\cal V}$ is a generic scalar operator with ghost number $0$
 and the last equality holds only on the unit circle.
  Note that $c^{\sigma}\left( \sigma\right)$, the tangent component of the
 ghost field along the boundary of the disc, is the only nonvanishing
component
on the boundary. See eq. (\ref{p=-1}) in the Appendix .
 It is safe to obtain the $BRS$ transformation
\beq
\label{deltabrs}
\delta_{BRS} {\cal O} =  i \epsilon  \{ Q_{BRS}, {\cal O} \} \;\;,
\eeq
 using a free open string:
\beq
Q_{BRS} = \frac{1}{2\pi i} \oint  dz j_{BRS}   \;\;.
\eeq
\bea
j_{BRS}(z)&=&
c^z(z)T^{ m}_{zz}(z)+
:b_{zz}(z)c^z(z) \partial_z c^z(z):
+\frac{3}{2} {\partial_z}^2 c^z(z)
\nonumber \\
T^{ m}_{zz}(z)&=&
-\frac{1}{\alpha'}
:\partial_z X^{\mu} \partial_z X_{\mu}:  \;\;\;.
\eea
We find
\beqn
  \{ Q_{BRS}, {\cal O}  \left(w, \bar{w} \right) \} =
   c^{\sigma} \partial_{\sigma} c^{\sigma} \left( w \right)
\left( \alpha^{\prime}
\frac{ \partial^{2}}{\partial X^{\mu} \partial X_{\mu}} +1 \right)
{\cal V}   \left( w, \bar{w}   \right) \;\;\;.
\eeqn
 We confirm that an on-shell open string tachyon with intercept $-1$
 is   represented by the vertex operator ${\cal V} = \exp ik \cdot X$
 with $\alpha^{\prime} M^{2} = - \alpha^{\prime} k^{\mu} k_{\mu} = -1$.

\subsection{ $dS^{disc} = i_{V} \omega$}

 The defining differential equation (\ref{sdomega}) for $S^{disc}$
 is written as
\beqn
 \frac{\delta S^{disc} }{\delta \lambda_{\alpha}}
 &=& \frac{K}{2} \int \int \frac{d\sigma}{2\pi}
   tr \frac{d\sigma^{\prime}}{2\pi}
\langle {\cal O}^{\alpha}\left( \sigma\right)
\{ Q_{BRS}, {\cal O} \}\left( \sigma^{\prime} \right)
 \rangle_{ {\cal V}, disc}^{ghost} \;\;\;, \\
{\cal O}\left( \sigma\right) &=&  c^{\sigma}\left( \sigma\right)
{\cal V}\left( \sigma\right)  = \sum_{\alpha}\lambda_{\alpha}
c^{\sigma}\left( \sigma \right) {\cal V}^{\alpha} \left( \sigma\right)
  = \sum_{\alpha}\lambda_{\alpha} {\cal O}^{\alpha}\left( \sigma\right)
\;\;.
\eeqn
Here
$\langle   \cdots \rangle_{ {\cal V}, disc}^{ghost}$
is the unnormalized path integral with respect to the worldsheet matter
action
\beqn
 I^{disc} &=& I^{bulk} +  \int_{\partial \Sigma}
 \frac{d\sigma}{2\pi} {\cal V}\left( \sigma\right)
\;\;,  \\
I^{bulk} &=&   \frac{1}{4\pi \alpha} \int_{\Sigma} \sqrt{\det \eta}
  \eta^{ij} \partial_{i}X^{\mu} \partial_{j}X^{\nu} g_{\mu \nu} d^{2} \sigma
\;\;.
\eeqn
 Here  $\eta^{ij}$ is a worldsheet metric which is taken to be flat, and
$g_{\mu \nu}$ is a spacetime metric with Minkowski signature.
 We denote by $tr$ the trace over the Chan-Paton space. This simply gives
 us a factor $n$ of $so(n)$ Lie algebra.
  Separating the ghost Hilbert space, we find
\beqn
 \frac{\delta S^{disc}}{\delta \lambda_{\alpha}}
 = K n \int \int \frac{d\sigma}{2\pi}
\frac{d\sigma^{\prime}}{2\pi}
  \left( 1- \cos \left( \sigma - \sigma^{\prime}  \right) \right)
 \langle c_{1} c_{0} c_{-1} \rangle_{disc}  \nonumber \\
\langle {\cal V}^{\alpha} \left( \sigma\right)  \left( \alpha^{\prime}
\frac{ \partial^{2}}{\partial X^{\mu} \partial X_{\mu}} +1 \right)
{\cal V} \left( \sigma^{\prime} \right)  \rangle_{ {\cal V}, disc} \;\;.
\label{diff-eq-disc}
\eeqn

\subsection{Off-shell Boundary State}
  The unnormalized matter path integral
 $\langle \cdots \rangle_{ {\cal V}, disc}$ can be
  represented  as a matrix element
\beq
\label{vev=B0}
 \langle   \cdots \rangle_{ {\cal V}, disc}  =  \langle  B \mid  \cdots
  \mid 0  \rangle_{ {\cal V},  disc} \;\;.
\eeq
  The bra vector  $\langle  B \mid$ obeys
\beq
\label{Bcondition}
 \langle  B \mid    \left. \left(  \frac{1}{2\pi \alpha^{\prime}}
 \left( z \frac{\partial}{\partial z} +
\bar{z}
 \frac{\partial}{\partial\bar{z} } \right)  X^{\mu} +   \frac{1}{2 \pi}
 \frac{\partial {\cal V}}{\partial X^{\mu}}  \right)
 \right|_{z=e^{i\sigma}, \; \bar{z}= e^{-i\sigma} }    =  0\;\;.
\eeq
  We refer to $\langle B \mid $ as an off-shell boundary state of
the disc.
 The condition (\ref{Bcondition}) is a consequence from
 the correspondence between the matrix element (\ref{vev=B0}) and the end point of  the path integral. The latter one obeys
 the boundary condition  derived from the worldsheet action $I^{disc}$
on the disc.
 Eq. (\ref{Bcondition}) tells us that,
  at the initial point of the coupling constant flow,  the system obeys the
Neumann  boundary condition:
\beq
\label{BNeumann}
 \langle  B \mid   \left( \alpha_{n} +  \tilde{\alpha}_{-n} \right) =0 \;\;.
\eeq
 The end point of the flow is described by zero of
 $\frac{\partial {\cal V}}{\partial X^{\mu}} $, namely, the Dirichlet
boundary condition:
\beq
\label{BDirichlet}
 \langle  B \mid \left(  \alpha_{n} -  \tilde{\alpha}_{-n} \right) =0 \;\;.
\eeq

 An explicit interpolation between the two ends can be done
 in solvable cases. In the case of quadratic profiles,
  the proper form of the Green function has been given in
 \cite{Lee,Fujii-Itoyama-2}.
 For analyses which utilize the boundary sine-Gordon model
  and the $g$ function $= \langle B \mid 0\rangle $, see
\cite{harvey-kutasov-Martinec,Fujii-Itoyama-1}.

\section{$RP^{2}$ Extension }
We now proceed to construct the
$RP^{2}$ extension of BSFT.
 $RP^{2}$ is a nonorientable Riemann surface of Euler number one
 with no hole, no boundary and one crosscap.
It is a geometry swept by a closed string.
The external physical states of this nonorientable closed string are
represented
 by the vertex operators inserted at any point on the
surface.
The physical state conditions are stated as
$\{Q_{BRS},  {\cal O}^{closed} \}=0$.
Here $Q_{ BRS}$ is the $BRS$ operator on $RP^2$
and ${\cal O}^{closed}$ is a generic vertex operator.
In the case of the ground state scalar of a closed unoriented bosonic
string,
this leads us to the on-shell condition
$\alpha' M^{2} = -\alpha' k_{\mu} k^{\mu} =-4$, namely the closed string
tachyon with
intercept $-4$.
For completeness of our discussion,
we will include its derivation in the Appendix B.

Another property of $RP^2$, which is of vital importance to us, is
${\bf \pi}_1(RP^2)={\bf Z}_2$.
This permits us to consider a nontrivial loop ${\cal C}$ which is
 represented by a
path connecting any two conjugate points (for example, $1$ and  $-1$)
on the complex plane, ${\it i.e.}$  the double of this Riemann surface.
As a reference loop, we consider   half of a unit circle
$z=e^{i\sigma}$ $(0\le \sigma < \pi)$.

 \subsection{Glossary of Notation}
Let us first summarize our notation in the $RP^2$ case.
(See Appendix for more detail.)
The tangential component and the normal component of the ghost field
$c^{\sigma}$ and $c^{r}$ along the loop are respectively
\begin{eqnarray}
c^{\sigma}
&\equiv&
\frac{\partial \sigma}{\partial z}c^{z}
+
\frac{\partial \sigma}{ \partial \bar{z} }c^{\bar{z}}
=
-\frac{i}{2}
\left(
\frac{c^{z}(z)}{z}
-\frac{ \bar{c}^{\bar{z}}(\bar{z}) }{\bar{z}}
\right)
=
-i \frac{ c^{z}(z)-c^{z}(-z) }{2z}\;\;,   \\
c^{r}
&\equiv&
\frac{\partial r}{\partial z}c^{z}
+
\frac{\partial r}{\partial \bar{z}}c^{\bar{z}}
=
\frac{1}{2r}
\Big(
{\bar z}c^{z}(z)+z\bar{c}^{\bar{z}}(\bar{z})
\Big)
=
\frac{c^{z}(z)+c^{z}(-z)}{2z}\;\;.
\end{eqnarray}
In both equations, the last equality holds only when $|z|=1$.
We also note
\begin{eqnarray}
c^{\sigma}\partial_{\sigma}c^{\sigma}(w)
&=&
-i \frac{c^{z}(w)-c^{z}(-w)}{2w}\:
\frac{\partial_z c^{z}(w)+(\partial_z c^{z})(-w)}{2},
\\
c^{\sigma}(w)\partial_{\sigma}{\cal V}(X^{\mu})
&=&
\frac{c^{z}(w)-c^{z}(-w)}{2}
\Big(
\partial_z X^{\mu}(w)+(\partial_z X^{\mu})(-w)
\Big)
\frac{\partial}{\partial X^{\mu}}{\cal V}.
\end{eqnarray}
Here the argument $-w$ should be substituted after the derivatives are taken
and ${\cal V}$ is an arbitrary function of $X^{\mu}$.

 \subsection{ BRS charge $Q^{(\sigma)}$ }
Our first objective is to obtain a BRS charge
which corresponds to the vector field $V$ in the BV formalism
for the $RP^2$ case.
Let us consider the following expression:
\begin{eqnarray}
Q^{(\sigma)}_{BRS}
&\equiv&
\oint
\frac{dz}{2\pi i}j^{BRS}_z (z)\;\;,
\nonumber \\
j^{BRS}_z (z)
&\equiv& 2
c^z_{\mbox{\scriptsize even}}(z)
\left( -\frac{1}{\alpha '}\right)
:(\partial_z X^{\mu})_{\mbox{\scriptsize even}}(z)
(\partial_z X_{\mu})_{\mbox{\scriptsize even}}(z):
\nonumber \\
& &+
:b_{zz\ \mbox{\scriptsize even}}(z)
c^z_{\mbox{\scriptsize even}}(z)
\partial_z c^z_{\mbox{\scriptsize even}}(z):
+ \mbox{total derivatives}\; ,
\label{maru2}
\end{eqnarray}
where  the subscript $\mbox{\scriptsize even}$ implies that
  the modes are restricted to the even ones. For example,
$c^z_{\mbox{\scriptsize even}}(z)\equiv
{\displaystyle \sum_{n\ \mbox{\scriptsize even}} } c_n z^{-n+1}
=\frac{c^z(z)-c^z(-z)}{2}$.
We will now show that our $Q^{(\sigma)}_{BRS}$,
when acted upon the conformal fields on $|z|=1$,
generates the BRS transformations representing the diffeomorphisms
in the $\sigma$ direction.

Let us first consider the ghost part of our BRS current:
\begin{eqnarray}
j_z^{(g)}(z)
\equiv
\frac{1}{8}
\Big(b_{zz}(z)+b_{zz}(-z)\Big)
\Big(c^{z}(z)-c^{z}(-z)\Big)
\Big(\partial_z c^{z}(z)+(\partial_z c^{z})(-z)\Big).
\label{maru3}
\end{eqnarray}
We obtain
\begin{eqnarray}
j_z^{(g)}(z) c^{\sigma}(w)
\sim
-\frac{i}{2}
\left(\frac{1}{z-w}-\frac{1}{z+w}\right)
\frac{c^{z}(z)- c^{z}(-z)}{2w}\:
\frac{\partial_z c^{z}(z)+(\partial_z c^{z})(-z)}{2},
\end{eqnarray}
so that
\begin{eqnarray}
\delta_{B} c^{\sigma}(w)
&\equiv&
i\epsilon
\left\{
\oint \frac{dz}{2\pi i} j_z^{(g)}(z),
c^{\sigma}(w)
\right\}
\nonumber \\
&=&
i\epsilon
c^{\sigma}\partial_{\sigma} c^{\sigma}(w).
\end{eqnarray}
One can easily show
$\delta_{B} c^{r}(w)=0$.
These are the desirable $BRS$ transformation laws for
 the $\sigma$ diffeomorphisms.
(Again $|w|=1$ is understood.)

Next, we consider the matter part of the BRS current:
\begin{eqnarray}
j_z^{(m)}(z)
\equiv 2
\frac{c^{z}(z)  - c^{z}(-z)}{2}
\left(-\frac{1}{\alpha'}\right)
: (\partial_z X^{\mu})_{\mbox{\scriptsize even}}(z)
(\partial_z X_{\mu})_{\mbox{\scriptsize even}}(z) : \;\;.
\label{maru4}
\end{eqnarray}
We find
\begin{eqnarray}
j_z^{(m)}(z) {\cal V}^{\prime}(X^{\mu}(w,\bar{w}))
\sim
\frac{c^{z}(z)-c^{z}(-z)}{2}
\left(\frac{1}{z-w}+\frac{1}{z+w}\right)
: (\partial_z X^{\mu})_{\mbox{\scriptsize even}}(z)
\frac{\partial}{\partial X^{\mu}}
{\cal V}^{\prime}(X^\mu):
\nonumber \\
-\frac{\alpha'}{8}
\frac{c^{z}(z)-c^{z}(-z)}{2}
\left(\frac{1}{(z-w)^2}+\frac{2}{(z-w)(z+w)}+\frac{1}{(z+w)^2}\right)
: \frac{\partial^2}{\partial X^{\mu} \partial X_{\mu}} {\cal
V}^{\prime}(X) :,
\mbox{        }
\end{eqnarray}
so that
\begin{eqnarray}
\left\{
\oint \frac{dz}{2\pi i} j_z^{(m)}(z),
{\cal V}^{\prime}(X^{\mu}(w,\bar w))
\right\}\!\!
=\!\!
c^{\sigma}\partial_{\sigma} {\cal V}^{\prime}(X^{\mu})\!
-\!
\frac{\alpha'}{4}
\!\!
\left(\partial_{\sigma}c^{\sigma}(w)\!\!+2ic^{\sigma}(w)\right)\!\!
:\!\!
\frac{\partial^2}{\partial X^{\mu} \partial X_{\mu}} {\cal
V}^{\prime}(X^{\mu})
\!\! :\!\!.
\mbox{                 }
\end{eqnarray}
In particular, setting
${\cal V}^{\prime}(X^{\mu})=X^{\mu}$, we obtain
\begin{eqnarray}
\delta_{B}X^{\mu}
\equiv
i\epsilon
\left\{
\oint \frac{dz}{2\pi i} j_z^{(m)}(z),
X^{\mu}
\right\}
=i\epsilon c^{\sigma}\partial_{\sigma} X^{\mu}(w)\;\;.
\end{eqnarray}
This is again the desirable BRS transformation.
%
Thus, the operator
\begin{eqnarray}
Q^{(\sigma)}_{BRS}=
\oint \frac{dz}{2\pi i}
\left( j_z^{(g)}(z)+j_z^{(m)}(z) \right) \;\;.
\end{eqnarray}
generates the BRS transformations $\delta_{B}$
associated with the $\sigma$ diffeomorphisms on
$|w|=1$.

\subsection {$\{Q^{(\sigma)}_{BRS},{\cal O}\}$ }
It is now immediate to carry out the action of the $BRS$ charge
on a generic operator with ghost number one.
This also brings us an operator which is "on-shell" with respect to
$Q^{(\sigma)}_{BRS}$,  that is,
invariant under $\delta_{B}$.
Let ${\cal O}$ be a scalar operator with ghost number one,
\begin{eqnarray}
{\cal O}(w,\bar{w})
=
c^{\sigma}(w){\cal V}^{ \prime \sigma}(X^{\mu}(w,\bar{w}))
+
c^{r}(w){\cal V}^{ \prime r}(X^{\mu}(w,\bar{w}))\;\;.
\end{eqnarray}
We find
\begin{eqnarray}
\left\{
Q^{(\sigma)}_{BRS},{\cal O}(w,\bar{w})
\right\}
&=&
c^{\sigma}\partial_{\sigma}c^{\sigma}
\left(
1+\frac{\alpha'}{4} \frac{\partial^2}{\partial X^2}
\right)
{\cal V}^{\prime \sigma}(X)
\nonumber \\
&-&
c^r c^{\sigma}
\left(
\partial_{\sigma} {\cal V}^{\prime r} (X)
 -i\frac{\alpha'}{2} \frac{\partial^2}{\partial X^2} {\cal V}^{\prime r} (X)
\right)
+
\frac{\alpha'}{4}c^r \partial_{\sigma}c^{\sigma}
\frac{\partial^2}{\partial X^2} {\cal V}^{\prime r} (X).
\mbox{        }
\end{eqnarray}
The right hand side vanishes when
\begin{eqnarray}
\label{nonlooponshell}
{\cal V}^{\prime \sigma}(X)
&=&
\beta^{\sigma} \exp\{ik\cdot X(w,\bar{w})\} \;\;,
\nonumber \\
\alpha' M^2 &\equiv&  -\alpha' k^2=-4 \;\;, \nonumber \\
{\cal V}^{\prime r}(X)
&=&
\beta^{r} \mathbf{1}\;\;.
\end{eqnarray}
Here $\beta^{\sigma}$ and $\beta^{r}$ are constants and $\mathbf{1}$
is the identity operator.
 Eq. (\ref{nonlooponshell}) includes the case
\beqn
\label{onshellspecialcase}
{\cal V}^{\prime \sigma}(X)
&=&
\beta^{\sigma} \exp\{ik\cdot X(w,\bar{w})\} \;\;,
\nonumber \\
\alpha' M^2 &\equiv&  -\alpha' k^2=-4 \;\;, \nonumber \\
{\cal V}^{\prime r}(X)
&=&  0\;\;,
\end{eqnarray}
 as a special case. In what follows, we will consider
   a off-shell deformation of the following kind:
\begin{eqnarray}
 {\cal O}=c^{\sigma} {\cal V}^{\prime \prime}  + c^{r} {\cal V}^{\prime}
\;\;,
\label{maru5}
\end{eqnarray}
 where ${\cal V}^{\prime}$, ${\cal V}^{\prime \prime}$  are generic scalar
  relevant operators.

 \subsection{Nilpotency of $\delta_{B}$ }
The condition we need in our formalism is
$\delta_{B}^2 {\cal O}=0$ with ${\cal O}$
given by eq. (\ref{maru5}).
Using $\delta_{B}^2 c^{\sigma}=0$,
we obtain
\begin{eqnarray}
\delta_{B}^2 {\cal O}
=
c^{\sigma}\delta_{B}^2 {\cal V}^{\prime \prime}
+ c^{r}\delta_{B}^2 {\cal V}^{\prime}.
\end{eqnarray}
On the other hand, for both
${\cal V}= {\cal V}^{\prime}, {\cal V}^{\prime \prime}$
\begin{eqnarray}
\delta_{B}^2 {\cal V}
&=&
i\epsilon \delta_{B}
\left\{
 c^{\sigma}\partial_{\sigma} {\cal V}
 -\frac{\alpha'}{4}
 \left(
   \partial_{\sigma} c^{\sigma}
   +2i c^{\sigma}
 \right)
 \frac{\partial^2}{\partial X^2} {\cal V}
\right\}
\nonumber \\
&=&
i\epsilon
\biggl\{
   (\delta_{B} c^{\sigma}) \partial_{\sigma} {\cal V}
   -c^{\sigma} \partial_{\sigma} (\delta_{B} {\cal V})
\nonumber \\
& &-\frac{\alpha'}{4}
   \left(
     \partial_{\sigma}(\delta_{B} c^{\sigma})
     +2i \delta_{B} c^{\sigma}
   \right)
   \frac{\partial^2}{\partial X^2} {\cal V}
\nonumber \\
& &+\frac{\alpha'}{4}
 \left(
   \partial_{\sigma} c^{\sigma}
   + 2i c^{\sigma}
 \right)
   \delta_{B}
   \frac{\partial^2}{\partial X^2} {\cal V}
\biggl\}.
\end{eqnarray}
  The terms in the right hand side  cancel with one another and
we establish $\delta_{B}^2 {\cal O}=0$.

\subsection{ $dS^{RP^2}$}
The analyses made and the properties established above
provide the ingredients necessary for us to define
$dS$ in the framework of BV on $\Sigma^{\prime} =RP^2$ as well.
  Following the disc case, we introduce
 a defining differential equation (\ref{sdomega}) for $S^{RP^2}$
\beqn
&&
\frac{\delta S^{RP^2} }{\delta \lambda_{\alpha}}
=
 \frac{K^{\prime}}{2} \int_{\cal C} \int_{\cal C}
 \frac{d\sigma}{2\pi} \frac{d\sigma^{\prime}}{2\pi}
 \langle {\cal O}^{\alpha}\left( \sigma\right)
 \{Q_{BRS}^{(\sigma)} , {\cal O} \}\left( \sigma^{\prime} \right)
 \rangle_{ {\cal V}', RP^2}^{ghost} \\
&=&
 \frac{K^{\prime}}{2} \int_{\cal C} \!\! \int_{\cal C}\!
 \frac{d\sigma}{2\pi} \frac{d\sigma^{\prime}}{2\pi}
  \bigg\langle
  c^r(\sigma) \left(c^r c^{\sigma}\right)(\sigma ')
 \bigg\rangle_{RP^2}
 \bigg\langle
   {\cal V}^{\prime \alpha}(\sigma) (-)
   \left(
     \partial_{\sigma'} {\cal V}^{\prime}(\sigma ')
       -i \frac{\alpha'}{2}
       \frac{\partial^2}{\partial X^{\mu} \partial X_{\mu}} {\cal
V}^{\prime}
(\sigma')
   \right)
  \bigg\rangle_{ {\cal V}', RP^2} \;\;  \nonumber \\
&=&
 K^{\prime} \!\! \int_{\cal C} \!\! \int_{\cal C} \!
 \frac{d\sigma}{2\pi} \frac{d\sigma^{\prime}}{2\pi}
  sin(\sigma-\sigma')
 \big\langle c_1 c_0 c_{-1} \big\rangle_{RP^2} \!\!\
 \bigg\langle
   {\cal V}^{\prime \alpha}(\sigma)
   \left(
     \partial_{\sigma'} {\cal V}^{\prime}(\sigma ')
       -i \frac{\alpha'}{2}
       \frac{\partial^2}{\partial X^{\mu} \partial X_{\mu}}
{\cal V}^{\prime}(\sigma')
   \right)\!\!
  \bigg\rangle_{ {\cal V}',  RP^2}\!\!\!\!,   \\
  &&  {\cal V}^{\prime} \equiv  \sum_{\alpha} \lambda_{\alpha}
   {\cal V}^{\prime \alpha}\;\;.
\label{diff-eq-RP2}
\eeqn
  Here  the unnormalized path integral
 $\bigg\langle \cdots  \bigg\rangle_{ {\cal V}'}$ is evaluated with respect
 to
\beqn
 I^{RP^2} &=& I^{\prime bulk} +  \int_{ {\cal C}}
 \frac{d\sigma}{2\pi} {\cal V}^{\prime} \left( \sigma\right)
\;\;,  \\
I^{\prime bulk} &=&   \frac{1}{4\pi \alpha'}
 \int_{\Sigma^{\prime} -{\cal C} } \sqrt{\det \eta}
  \eta^{ij} \partial_{i}X^{\mu} \partial_{j}X^{\nu} g_{\mu \nu} d^{2} \sigma
\;\;.
\eeqn
Note that, in eq. (\ref{diff-eq-RP2}), the allowed form of
 the  nonvanishing ghost three point function has selected ${\cal
V}^{\prime}$
 alone and  ${\cal V}^{\prime \prime}$ has disappeared.

\subsection{The Off-shell Crosscap States}

 The unnormalized path integral on $RP^{2}$  geometry can be represented as
 a matrix element between  the ket vector of the closed string vacuum  and
  the bra vector of the off-shell crosscap state $\langle  C \mid$:
\beq
\label{vev=C0}
 \langle   \cdots \rangle_{ {\cal V}^{\prime}}  =  \langle  C \mid  \cdots
  \mid 0  \rangle_{ {\cal V}^{\prime}} \;\;.
\eeq
 The derivation of the equation which the bra vector
  $\langle  C \mid$ obeys
 involves  a closed  and
  unoriented nature of the Riemann surface $RP^{2}$.
 The fundamental domain can be represented by
  $\Sigma^{\prime} = \{ \{ r < 1 \}\} \cup \{ \{ r=1, 0 \leq \sigma <
\pi \} \}$ on the
complex plane.
 The variation of the action $I^{RP^2}$ must be carried out
consistently with
   the antipodal identification $ P(z) = - \frac{1}{\bar{z}}$.
 We find
\beqn
   & & \delta  I^{RP^2} =     \nonumber \\
  & &  \left.
 \frac{1}{2\pi \alpha^{\prime}}
 \int_{0}^{\pi}d \sigma  \delta X^{\mu} (w \frac{d}{dw} +
 \bar{w} \frac{d}{d\bar{w}} ) X^{\mu}\left(w, \bar{w} \right)
 \right|_{w=z, \; \bar{w}=  \bar{z} }     +
  \left.   \int_{0}^{\pi}d \sigma  \delta X^{\mu} \frac{1}{2\pi}
 \frac{\partial {\cal V}^{\prime} (X^{\mu}\left(w, \bar{w} \right) )}{\partial X^{\mu}}
 \right|_{w=z, \; \bar{w}= \bar{z} } \;\;  \nonumber \\
   & &  \left.
 \frac{1}{2\pi \alpha^{\prime}}
 \int_{0}^{\pi}d \sigma  \delta X^{\mu} (w \frac{d}{dw} +
 \bar{w} \frac{d}{d\bar{w}} ) X^{\mu}\left(w, \bar{w} \right)
 \right|_{w= -1/\bar{z}, \; \bar{w}= -1/z }     +
  \left.   \int_{0}^{\pi}d \sigma  \delta X^{\mu} \frac{1}{2\pi}
 \frac{\partial {\cal V}^{\prime} (X^{\mu}\left(w, \bar{w} \right) )}{\partial X^{\mu}}
 \right|_{w= -1/\bar{z}, \; 
\bar{w}= -1/z } \;\;  \nonumber \\
 &+&  {\rm the \; part \; proportional \; to \; eq. \; of \;
 motion}\;\;.
\eeqn
  Here, $z=(1-\delta)e^{i\sigma}$ and $\delta$ is a small but
 nonvanishing number. Reflecting the fact that
$RP^{2}$ is nonorientable,  we have nonvanishing contributions  from
 an open cover of ${\cal C}= \{ \{ r=1, 0 \leq \sigma < \pi \} \}$, namely,
  the edge of the fundamental domain $\Sigma^{\prime}$ despite that there is no boundary.
  Stokes theorem does not apply.

   From the correspondence between the matrix element and
  the end point condition of the path integral,
we conclude that the bra vector  $\langle  C \mid$ must obey
\beqn
\label{Ccondition}
 \langle  C \mid  K_{\mu}( z, \bar{z}; \delta) =0\;\;,  \nonumber\\
K_{\mu}( z, \bar{z}; \delta) \equiv
   \left. \left(  \frac{1}{2\pi \alpha^{\prime}}
 \left( w \frac{\partial}{\partial w} +\bar{w}
 \frac{\partial}{\partial\bar{w} } \right)  X^{\mu} +   \frac{1}{2 \pi}
 \frac{\partial {\cal V}^{\prime}}{\partial X^{\mu}}  \right)
 \right|_{w=z, \; \bar{w}= \bar{z} }   \nonumber \\
   +  \left. \left(  \frac{1}{2\pi \alpha^{\prime}}
 \left( w \frac{\partial}{\partial w} +\bar{w}
 \frac{\partial}{\partial\bar{w} } \right)  X^{\mu} +   \frac{1}{2 \pi}
 \frac{\partial {\cal V}^{\prime}}{\partial X^{\mu}}  \right)
 \right|_{w= -1/\bar{z}, \; 
\bar{w}= -1/z } \;\;.
\eeqn
 Expansion of this condition in $\delta$ generates conditions
 which may be referred to as off-shell crosscap conditions.
The initial point of the coupling constant flow is described as
  the ordinary crosscap condition:
\beq
\label{CC}
  \langle  C \mid    \left( \alpha_{n} +  (-)^{n} \tilde{\alpha}_{-n}
\right) =0 \;\;.
\eeq
 The end point of the flow is described by zero of
 $\frac{\partial {\cal V}^{\prime}}{\partial X^{\mu}} $, which is
 nothing but the orientifold condition.
\beq
\label{CCTdual}
  \langle  C \mid  \left( \alpha_{n} - (-)^{n} \tilde{\alpha}_{-n}
\right) =0 \;\;.
\eeq
 We have derived the two point function from eq. (\ref{Ccondition})
 in the case that ${\cal V}^{\prime}$ is quadratic in $X^{\mu}$
 and checked that it takes the proper form on both ends of the
 flow \footnote{ These will appear elsewhere together with
  other related issues.}.

 Let us finish this section by discussing 
  the properties of orientifold planes in general. They are
 certainly nondynamical objects in string perturbation theory.
 Nonperturbatively, they may become dynamical however. 
It is known in fact that the orientifold planes do become 
dynamical in Sen's
  description of F theory \cite{Sen} which connects  F 
 to orientifolds and the Seiberg-Witten curve of the UV finite
  4 dimensional N=2 supersymmetric gauge theory.
  While we are not able to tell the nonperturbative dynamical
 origin  which causes
 orientifold planes to get deformed, our construction does provide
 a proper description once this happens. 

\section{Total Action }

  Our construction of the action for open-closed string field theory is
 completed by adding $S^{disc}$ and $S^{RP^2}$.
\beq
 S = S^{disc} + S^{RP^2} \;\;.
\eeq
 The individual parts  $S^{disc}$ and $S^{RP^2}$ obey the first order
 differential equations   eqs. (\ref{diff-eq-disc}) and (\ref{diff-eq-RP2})
respectively.
  Here we discuss  the  normalization constants  appearing in these two
quantities.
 It has been shown in \cite{wittenbsft,shatashvili} that $S^{disc}$
on-shell,
(that is, with no operator insertion) is actually the matter partition
 function of a free open string up to an overall normalization.
 The same applies to $S^{RP^2}$ as well: on-shell
 it is the $RP^2$ matter partition function up to an overall normalization.
 We write these relations as
\beq
\label{sdisczdisc}
  \left.  S^{disc} \right|_{on-shell} = Z^{disc}/ g_{disc} \;\;.
\eeq
\beq
\label{srp2zrp2}
  \left.  S^{RP^2} \right|_{on-shell} = Z^{RP^2}/ g_{RP^2}   \;\;.
\eeq
  Here $Z^{disc}$ and $Z^{RP^2}$ are complete partition functions
  obtained by carrying out the path integrals over $X^{\mu}$ and the
two-dimensional metric.
(They are actually free(vacuum) energies of a string as a single string
can produce
  only a connected graph.)
  We denote by $g_{disc}$ and $g_{RP^2}$  the contributions  in
respective geometries from
  the path integrals of the two dimensional metric in the Polyakov
  formulation.  They are made of the ghost determinant divided by the
volume of the
  conformal killing vector and the order of the disconnected
diffeomorphisms.

Eq. (\ref{sdisczdisc}) and eq. (\ref{srp2zrp2}) can be related
  by invoking the cancellation between the dilaton
tadpole of disc and that of $RP^2$
\cite{Grins-Wise,Doug-Grins,Itoyama-Moxhay}.
 The operator insertion of a zero momentum tadpole in the first
quantized string can
 be done  by taking a derivative of the partition function
  with respect to the tension of a fundamental string
  $T^{fund} = \frac{1}{2\pi \alpha'}$.

 On the other hand, the BV formalism guarantees that the right hand side of
 the basic equation $dS = i_{V} \omega$ is a total differential at least
locally. The integration
  constant is, therefore, an additive one and, up to this constant,
 $S^{disc/RP^2}$ is completely determined:
\beqn
  S^{disc} &=& K n
    f^{disc} \left( \lambda^{\alpha} \right)
    \nonumber \\
 S^{RP^2}  &=& K^{\prime}
    f^{RP^2} \left( \lambda^{\alpha} \right)\;\;,
\eeqn
 where  $f^{disc} \left( \lambda^{\alpha} \right)$
 and  $f^{RP^2} \left( \lambda^{\alpha} \right)$ are both scalar functions,
  and can be determined by using solvable tachyonic profiles\footnote{
  We have set $\langle c_{1}c_{0}c_{-1} \rangle_{disc}
 =\langle c_{1}c_{0}c_{-1} \rangle_{RP^2} =1$. }.

  Putting all these discussions together, we find that the cancellation of
  the dilaton tadpoles implies
\beq
\label{tadcan}
  \frac{\partial}{\partial T^{fund}}  \left. \left(
       K n  g_{disc}   f^{disc}   +
    K^{\prime} g_{RP^2}   f^{RP^2}    \right) \right|_{on-shell}  =0 \;\;.
\eeq
  In \cite{Gosen}, $K$ has been determined by
 demanding  that  a proper amplitude of a nearly on-shell three point
  open string tachyon be obtained.
There is no argument in the $RP^2$ case which
  parallels the disc case as closed string tachyon with intercept $-4$
  lives in the bulk of $RP^2$ and does not live in  our nontrivial closed
loop alone.
 We see that eq. (\ref{tadcan}) determines the normalization $K^{\prime}$
 instead.

A final remark of this subsection is on unoriented nature of
 our open and closed strings.
The open-closed string field theory constructed above is nonorientable
as our construction involves the nonorientable surface $RP^2$.
 On-shell, the state space
  of an unoriented open string is obtained by
 choosing ${\cal O}$
 vertex operators and  Chan-Paton indices such that the product
 is even under the twist operation $\Omega$. (We have nothing
 to say in this paper on the state space of an unoriented closed string.)
  Let us note, however, that $\Omega$ is a symmetry
 only at the two ends of the coupling constant flow.
 In the closed string picture, this is easily seen at
 our construction of the off-shell
  boundary/crosscap states.  Eqs. (\ref{BNeumann}), (\ref{BDirichlet}),
 (\ref{CC}), (\ref{CCTdual}) stay invariant under
  the twist operation as
\beqn
  \Omega  \left( \alpha_{n} \pm  \tilde{\alpha}_{-n} \right)  \Omega^{-1}
 =   \pm \left( \alpha_{-n} \pm  \tilde{\alpha}_{n} \right) \;\;,
\nonumber \\
   \Omega  \left( \alpha_{n} \pm  (-)^{n} \tilde{\alpha}_{-n} \right)
\Omega^{-1}
 =   \pm \left( \alpha_{-n} \pm  (-)^{n} \tilde{\alpha}_{n} \right) \;\;.
\eeqn
 At a generic value of the coupling,  neither of  eqs. (\ref{Bcondition}),
(\ref{Ccondition})  stays invariant.

\section{Acknowledgements}
  We thank Akira Fujii, Nobuyuki Ishibashi, Koichi Murakami and Toshio
Nakatsu
  for  useful discussion  on this subject.
 We are grateful  to  Santo seminar  for providing an opportunity
  for collaboration.

\newpage

\appendix

\setcounter{equation}{0}

\section{A}
 In this appendix, the mode expansion of a conformal field with
weight $p$ is obtained both on a unit disc $D_2$ and on
$RP^2$.
$D_2$ and  $RP^2$ are constructed from the complex plane by the
respective identifications
$z'=\frac{1}{\bar{z}}$ and $z'=-\frac{1}{\bar{z}}$.

Let $\Phi(z) \equiv
\Phi_{\underbrace{\scriptstyle z\cdots z}_{p}}
^{\overbrace{\scriptstyle z\cdots z}^{q}}(z)$
be a holomorphic field with
weight $p-q$ and $\bar{\Phi}(\bar{z})$ be an antiholomorphic field
 with weight $p-q$.
These mode expansions read respectively
\bea
\Phi(z)=\sum_{n \in Z} \phi_n z^{-n-p+q},
\:\:\:\:\:\:
\bar{\Phi}(\bar{z})=\sum_{n \in Z} \bar{\phi_n} \bar{z}^{-n-p+q}.
\eea
The modes of $\Phi(z)$ and those of $\bar{\Phi}(\bar{z})$ are
related to each other both on $D_2$ and  $RP^2$ and this relation can be
made
clear by extending the domain of $\Phi(z')$ to
$|z'| > 1$ through the respective involutions
$z'=\frac{1}{\bar{z}}$ ($D_2$) and
$z'=-\frac{1}{\bar{z}}$ ($RP^2$):
\bea
\bar{\Phi}(\bar{z})
={\left(\frac{\partial z'}{\partial \bar{z} }\right)}^{p-q}
\Phi(z').
\eea
On $|z|=1$, this leads to a relation
\bea
\left.(\bar{z})^{p-q} \bar{\Phi}(\bar{z})
\right|_{|z|=1}
=
\left.(\mp z)^{p-q} \Phi(\pm z)\right|_{|z|=1}.
\eea
The upper  and lower signs   refer respectively to the $D_2$ and $RP^2$
cases.
In terms of the modes,  this equation reads
\bea
\bar{\phi}_n =\left\{
\begin{array}{ll}
\phi_{-n} e^{-i(p-q)\pi}, & D_2 \\
\phi_{-n} e^{-i(p-q)\pi} (-)^n, & RP^2.
\end{array}
\right.
\eea
The radial, tangential and mixed components are constructed as
\beqn
\Phi_{\underbrace{\scriptstyle r \cdots r}_{p_1}
\underbrace{\scriptstyle \sigma \cdots \sigma}_{p_2} }
^{\overbrace{\scriptstyle r \cdots r}^{q_1}
\overbrace{\scriptstyle \sigma \cdots \sigma}^{q_2} }
(z,\bar{z})
    &\equiv&
\left(\frac{\partial z}{\partial r}\right)^{p_1}
\left(\frac{\partial z}{\partial \sigma}\right)^{p_2}
\left(\frac{\partial r}{\partial z}\right)^{q_1}
\left(\frac{\partial \sigma}{\partial z}\right)^{q_2}
\Phi(z)
\nonumber \\
& &+
\left(\frac{\partial \bar{z}}{\partial r}\right)^{p_1}
\left(\frac{\partial \bar{z}}{\partial \sigma}\right)^{p_2}
\left(\frac{\partial r}{\partial \bar{z}}\right)^{q_1}
\left(\frac{\partial \sigma}{\partial \bar{z}}\right)^{q_2}
\bar{\Phi}(\bar{z})   \\
  &=&
   \left.
  \left(\frac{z}{r} \right)^{p_1} (iz)^{p_2}
  \left(\frac{\bar{z}}{2r} \right)^{q_1}
  \left(\frac{1}{2iz} \right)^{q_2} \Phi(z)
\right.
\nonumber \\
& &+
\left.
 \left( \frac{\bar{z}}{r} \right)^{p_1} (-i\bar{z})^{p_2}
  {\left(\frac{z}{2r} \right)}^{q_1}
  {\left(-\frac{1}{2i \bar{z}} \right)}^{q_2} \bar{\Phi}(\bar{z})
\right.  \;\;.
\eeqn
On $|z|=1$,  we obtain
\bea
\left.
\Phi_{\underbrace{\scriptstyle r \cdots r}_{p_1}
\underbrace{\scriptstyle \sigma \cdots \sigma}_{p_2} }
^{\overbrace{\scriptstyle r \cdots r}^{q_1}
\overbrace{\scriptstyle \sigma \cdots \sigma}^{q_2} }
(z,\bar{z})
\right|_{|z|=1}
=  (i)^{p_2-q_2} \frac{1}{2^q} z^{p-q}
\left(
\left.  \Phi(z) \right|_{|z|=1}
+ (-)^{p_2-q_2} (\mp)^{p-q}
\left.  \Phi(\pm z) \right|_{|z|=1}
\right)  \;\;.
\eea
For example,
\begin{itemize}
\item{$p=1,q=0$:}
\bea
\alpha_z(z)&\equiv&
i\sqrt{\frac{2}{\alpha'}} \partial_z X(z,\bar{z})
=
\sum_n \alpha_n z^{-n-1}  \;\; ,\nonumber \\
\bar{\alpha}_{\bar{z}}(\bar{z})&\equiv&
i\sqrt{\frac{2}{\alpha'}}
\bar{\partial}_{\bar{z}} X(z,\bar{z})
=
\sum_n \tilde{\alpha}_n \bar{z}^{-n-1}  \;\;, \nonumber \\
 \alpha_{r}(z,\bar{z}) &=&  \frac{z}{r} \alpha_z(z) +
  \frac{\bar{z}}{r} \bar{\alpha}_{\bar{z}}(\bar{z})     \;\;,  \nonumber\\
 \alpha_{\sigma}(z,\bar{z})
 &=&  iz \alpha_z(z)  -i \bar{z} \bar{\alpha}_{\bar{z}}(\bar{z}) \;\;,
\nonumber\\
\left.
 \alpha_{r}(z,\bar{z})
\right|_{|z|=1}
&=&
\left\{
\begin{array}{ll}
0, & D_2 \\
\left.
z(\alpha_z (z)+\alpha_z (-z))
\right|_{|z|=1}, & RP^2
\end{array}
\right.   \;\;,
\nonumber \\
\left.
 \alpha_{\sigma}(z,\bar{z})
\right|_{|z|=1}
&=&
\left\{
\begin{array}{ll}
\left.
2iz \alpha_z (z)
\right|_{|z|=1}, & D_2 \\
\left.
iz(\alpha_z (z)-\alpha_z (-z))
\right|_{|z|=1}, & RP^2
\end{array}
\right.  \;\;,
\nonumber \\
\label{p=1}
\eea

\item{$p=0,q=1$:}
\bea
c^z(z)&=&\sum_n c_n z^{-n+1}  \;\;,
\:\:\:\:\:\:
\bar{c}^{\bar{z}}(\bar{z})= \sum_n \bar{c}_n \bar{z}^{-n+1} \;\;,
\nonumber \\
c^{r}(z,\bar{z})
&=&
\frac{1}{2r}
\Big(
{\bar z}c^{z}(z)+z\bar{c}^{\bar{z}}(\bar{z})
\Big)
\nonumber \\
c^{\sigma}(z,\bar{z})
&=&
-\frac{i}{2}
\left(
\frac{c^{z}(z)}{z}
-\frac{ \bar{c}^{\bar{z}}(\bar{z}) }{\bar{z}}
\right)
\nonumber \\
\left.
c^r(\sigma)\equiv c^r(z,\bar{z})
\right|_{|z|=1}
&=& \left\{
\begin{array}{ll}
0, & D_2 \\
\frac{1}{2z}
\left.
\left( c^z(z)+c^z(e^{i\pi}z)\right)
\right|_{|z|=1}
=  \displaystyle{ \sum_{n \: \mbox{\scriptsize odd}} }
c_n e^{-in\sigma}, & RP^2
\end{array}
\right.
\nonumber \\
 \left. c^{\sigma}(\sigma)\equiv c^{\sigma}(z,\bar{z})
\right|_{|z|=1}
&=&\left\{
\begin{array}{ll}
\left.
 \frac{1}{iz} c^z(z)
\right|_{|z|=1}
=-i \displaystyle{ \sum_n } c_n e^{-in\sigma},
& D_2 \\
\frac{1}{2iz}
\left.
\left( c^z(z)-c^z(e^{i\pi}z)\right)
\right|_{|z|=1}
= -i   \displaystyle{ \sum_{n \: {\mbox{\scriptsize even}}} }
c_n e^{-in\sigma}, & RP^2
\end{array}
\right.
\nonumber \\
\label{p=-1}
\eea

\end{itemize}
 Some of these formulas are exploited in the text.

\section{B}

Let ${\cal O}(z,\bar{z})$ be a generic scalar operator with
ghost number 2 on $RP^2$ geometry.
We  write this as
\begin{eqnarray}
{\cal O}(z,\bar{z})
=
\frac{c^z (z)}{z} \frac{\bar{c}^{(\bar{z})} (\bar{z})}{\bar{z}}
{\cal V}(z,\bar{z}) \;\;.
\end{eqnarray}
Let $Q_{BRS}$ be the BRS charge
obtained  from the integration of the holomorphic BRS current.
We find
\begin{eqnarray}
[ Q_{BRS}, {\cal O}(z,\bar{z}) ]
=
-
\Big(
\partial_z c^z (z) + (\partial_z c^z)(-1/\bar{z})
\Big)
\left(
1+\frac{\alpha'}{4} \frac{\partial^2}{\partial X^{\mu} \partial X_{\mu}}
\right)
{\cal O}(z,\bar{z})\;.
\end{eqnarray}
The on-shell ground state scalar is represented by the vertex operator
\beqn
{\cal V}(z,\bar{z})=e^{ik\cdot X(z,\bar{z})} \;\;, \\
  \alpha' M^2= -\alpha' k^2= -4 \;\;.
\eeqn
This latter condition is that of the tachyon of a closed unoriented
bosonic string with intercept $-4$.

\newpage


\end{document}